\documentstyle[12pt,aaspp4]{article}
\begin{document}
\tightenlines
%
%
\newcommand{\ea}{{et~al.}}
\newcommand{\IUE}{{\it IUE}}
\newcommand{\dmod}{$(m - M)_{0}$}
\newcommand{\logg}{$\log g$}
\newcommand{\logl}{$\log L$}
\newcommand{\lya}{\mbox{Ly$\alpha$}}
\newcommand{\lsun}{L$_{\sun}$}
\newcommand{\msun}{M$_{\sun}$}
\newcommand{\mv}{{M}$_{V}$ }
\newcommand{\rsun}{R$_{\sun}$}
\newcommand{\teff}{T$_{\rm eff}$}
\newcommand{\vsini}{$v \sin i$}
\newcommand{\rd}{Di\thinspace Stefano}
\newcommand{\re}{Einstein radius}
\newcommand{\er}{Einstein radius}
\newcommand{\ml}{microlensing}
\newcommand{\pl}{planet}
\newcommand{\mage}{magnification}
\newcommand{\lc}{light curve}
\newcommand{\ev}{event}
\newcommand{\ec}{encounter}
\newcommand{\pop}{population}
\newcommand{\mo}{monitoring}
\newcommand{\bi}{binary}
\newcommand{\bis}{binaries}
\newcommand{\ob}{observation}
\newcommand{\ps}{planetary system}
\newcommand{\sy}{system}
\newcommand{\pr}{program}
\newcommand{\res}{resonant}
\newcommand{\ov}{overlap}
\newcommand{\cs}{central star}
\newcommand{\inm}{innermost planet}
\newcommand{\dbn}{distribution}
\newcommand{\shdn}{short-duration}
\newcommand{\evd}{evidence}
\newcommand{\zres}{zone for resonant lensing}
\newcommand{\sep}{separation}
\newcommand{\dtn}{detection}
\newcommand{\bl}{blending} 
\newcommand{\mt}{monitoring teams} 
\newcommand{\fut}{follow-up teams} 
\newcommand{\ft}{follow-up teams} 
\newcommand{\fss}{finite-source-size} 
\newcommand{\fsse}{finite-source-size effects} 
\newcommand{\cc}{caustic crossing}
\newcommand{\Asun}{A$_{\sun}$}
\newcommand{\Ajup}{A$_J$}
\newcommand{\Anep}{A$_N$}
\newcommand{\Nsun}{N$_{\sun}$}
\newcommand{\goesto}{\longrightarrow}
\def\mr{multiple repetitions}
\def\op{orbital plane} \def\lp{lens plane}
\def\wo{wide orbit}
\def\otn{orientation} \def\vy{velocity} \def\vt{$v_t$}
\def\smn{simulation}
\def\em{Earth-mass}
\def\jm{Jupiter-mass}


\def\stacksymbols #1#2#3#4{\def\theguybelow{#2}
    \def\verticalposition{\lower#3pt}
    \def\spacingwithinsymbol{\baselineskip0pt\lineskip#4pt}
    \mathrel{\mathpalette\intermediary#1}}
\def\intermediary#1#2{\verticalposition\vbox{\spacingwithinsymbol
      \everycr={}\tabskip0pt
      \halign{$\mathsurround0pt#1\hfil##\hfil$\crcr#2\crcr
               \theguybelow\crcr}}}

\def\lapproxeq{\stacksymbols{<}{\sim}{2.5}{.2}}
\def\gapproxeq{\stacksymbols{>}{\sim}{3}{.5}}
\def\du{duration} \def\dca{distance of closest approach}
\def\stl{stellar-lens}
\vskip -.4 true in
\title{
Microlensing and the Search for Extraterrestrial Life
}

\author{Rosanne \rd\altaffilmark{1}} 

\altaffiltext{1}{
Harvard-Smithsonian Center for Astrophysics,
60 Garden St., Cambridge, MA 02138; e-mail:  rdistefano@cfa.harvard.edu}

\vspace{-0.15in}

\begin{abstract}

\vspace{-0.15in}
Are \ml\ searches likely to discover \pl s that harbor life?
Given our present state of knowledge, this is a difficult question
to answer. We therefore begin by asking a more narrowly focused question:
are conditions on \pl s discovered via \ml\ likely to be similar to
those we experience on Earth?
In this paper I link the \ml\ observations to the well-known ``Goldilocks
Problem" (conditions on the Earth-like \pl s need to be {\it ``just right"}),
to find that Earth-like \pl s discovered via \ml\ are likely to
be orbiting stars more luminous than the sun. This means that light
from the \ps 's  \cs\ may contribute a significant fraction of the
baseline flux relative to the star that is lensed. Such blending
of light from the lens with light from the lensed source
can, in principle, limit our ability to detect these \ev s.
This turns out not to be a significant problem, however. A second
consequence of blending is the opportunity to determine the
spectral type of the lensed star. This circumstance, plus the
possibility that \fsse\ are important, implies that some meaningful
follow-up observations are likely to  be
possible for a subset Earth-like \pl s
discovered via \ml . In addition, 
 calculations indicate that reasonable requirements on the
\pl 's density and surface gravity imply that the mass of Earth-like
\pl s is likely to be within a factor of $\sim 15$ of an Earth mass.

\end{abstract}
\vskip -.2 true in

\section{Microlensing and the Search for Extraterrestrial Life}

Recent and ongoing advances in technology have led to the discovery of
extrasolar planets
(e.g., Mayor \& Queloz 1995,
 Marcy {\it et el.} 1997a,b,
Butler \& Marcy 1996, Marcy \& Butler 1996,    
Cochran {\it et al.} 1997, Noyes {\it et al.} 1997;   
see also the references listed in 
the Encyclopedia of Extrasolar Planets,
www.obspm.fr/darc/planets/encycl.html),  
 and promise the discovery and even the imaging of
additional planets (Angel \& Woolf 1997; Fraclas \& Shelton 1997; Labeyrie
1996; Brown 1996).  These developments excite the imagination because
they seem to bring us closer to the possible discovery of extraterrestrial
life.  It is therefore interesting to ask whether microlensing searches are
likely to find planets on which life could thrive.

We do not yet have a clear enough understanding of the nature of life to
definitively answer this question, because the range of physical conditions 
compatible with life may well be wider than our limited experience would
at first suggest.  It has been proposed, for example,
that life
may exist on the outer planets of our own solar
system and/or on their moons. (See, e.g., 
Reynolds et al. 1983; 
Raulin {\it et al.} 1992; 
Sagan, Thompson,  \& Khare 1992;  
Williams, Kasting, \& Wade 1997, McCord {\it et al.} 1997.)  
It has even been postulated that life may exist in non-planetary
environments, including the interiors of stars and molecular clouds (see,
e.g., Feinberg \& Shapiro 1980).
It therefore makes sense to consider any \pl , however close to or far from its star, and whatever
the nature of the star, as a possible harbor for life.

Nevertheless,
in the absence of real information on the existence of life away from our
own planet, one question is clearly interesting:  will microlensing find
evidence of planets similar to Earth?  We must of course define what we mean
by ``similar to Earth".  If we mean that there is a chance that chemical
processes necessary for Earth-like life could occur, then we want to
consider planets which can have similar surface and atmospheric make-up, and
similar amounts of energy available to fuel the necessary chemical processes.
The range of planetary masses and distances from a solar-type star 
compatible with Earth-like conditions, particularly the presence of
liquid water, has been dubbed the ``Goldilocks Problem", and has
been studied by many researchers. (See, e.g., Rampino \& Caldeira 1994,
for an overview.) 

In linking the Goldilocks problem to \ml\ observations, we must focus on the 
properties of \ps s that determine their detectability when searched
for by \ml\ monitoring programs. \S 2 is therefore devoted to a brief overview of
the detection of \pl s via \ml . The question of whether planets 
discovered through their action as microlenses may be Earth-like,
is addressed 
in \S 3. 
One of the primary findings of this study
 is that when a planet discovered via \ml\
receives from its \cs\ a flux of radiation comparable to that received by the 
Earth from the Sun, the \cs\ will generally be more luminous than our Sun.
This means that light from the lensed star
 will be blended with light from the \ps . 
\S 4 is devoted to studying how blending (a) influences the
detection of, and (b) helps us develop strategies to study \ps s 
discovered through \ml. In \S 5, I 
summarize the conclusions.
             
\section{Microlensing and Planet Detection}

The detection of a \pl\  via \ml\ is possible if the separation between the
star and \pl\ is close to or larger than the Einstein radius, $R_E,$ of the
star: 
\begin{equation}
R_E = 9.0\, au\, \sqrt{\Big( {{M}\over{M_\odot}}\Big)\, \Big({{D_S}\over{10\, kpc}} \Big)\, x\, (1-x)},
\end{equation}
where $M$ is the stellar mass, $D_S$ is the distance to the lensed source,
and $x$ is the ratio of the distance to the lens to $D_S$. 
Consider lensing by a mass, $m_i$. Define $\tau_{E,i}$
to be the time taken for the track of a lensed star to cross a
distance in the lens plane equal to the Einstein diameter, $2\, R_{E,i}$. 
Note that the duration of the observable \ev\ generally differs from $\tau_{E,i}$ for 
several reasons. First, if the photometric sensitivity is good, 
the \ev\ may be detectable well before the track of the 
source comes within $R_E$ of the lens position. 
For example, the deviation from baseline is at the 
$\sim 6\%$ level, when the projected separation between the source and lens is
$2\, R_E.$ 
Second, the track of the source will not generally intersect
the lens position. Finally, blending may decrease 
and/or finite-source size effects may
increase 
the time during which the \ev\ is detectable.   
(See \rd\ 1998; \rd\ \& Scalzo 1997 or 1998a  for discussions relevant to
\pl\ lenses.)

Express the  
separation between the planet and star as $a= \mu\, R_E$. 
The minimum possible value of $\mu$ for which \pl s can be discovered via
\ml\ is $\sim 0.8$ (Gould \& Loeb 1992, but see
Wambsganss 1997 for a detailed discussion). 
For planets located between roughly $0.8\, R_E$ and $1.5\, R_E$
(the {\it \zres}),
the signature of a lensing \pl\ is a short-lived perturbation superposed
on a light curve whose underlying structure is 
due to lensing by the \ps 's star. The time duration of the
underlying event may be weeks or months. 
All of the literature published to date on planet discovery via \ml\ has been
limited to the possible discovery of \pl s in the \zres , and observing
programs designed to increase the detection efficiencies for
these types of \ev s are already underway. 
For a more detailed description
of these short-duration, ``resonant" perturbations, and
a discussion of the boundaries of \zres\ see, e.g., Mao \& Paczy\'nski 1991,
Gould \& Loeb 1992, 
Bennett \& Rhie 1996, Paczy\'nski 1996, Wambsganss 1997, Peale 1997.  

The benefits
of systematically extending the search, to look
 for \pl s located farther from the \cs ,
have recently begun to be explored (\rd\ \& Scalzo 1997, 1998a,b, \rd\ 1998).
For larger values of $a$, 
the \pl\ generally acts as an independent lens. When the \pl\ is the
only lens, then
the \ev\ will be an isolated \ev\ of \shdn. (For a Jupiter-mass \pl , e.g.,
the time duration of the perturbation from baseline lasts $\sim 3\%$ of the time
the deviation due to a solar-mass star would.)  
When the track of the source passes through the lensing region of one
\pl\ and also that of another system mass (typically the central star),
the event will appear to repeat (\rd\ \& Mao 1996, 
\rd\ \& Scalzo 1997, 1998a,b).  

There is no maximum value of $\mu$, other than that dictated by the
dynamics of the \ps\ itself--i.e., \pl s located too far from the \cs\
may be lost from the system.
Because the probability of
a repeating \ev\ falls off as $1/a$, where $a$ is the orbital separation,
while the probability of isolated \shdn\ \ev s is nearly independent of
position (and actually increases at the expense of repeating
\ev s as the orbital separation increases), 
isolated \shdn\ \ev s become the dominant mode of detection
for \pl s in wider orbits--particularly since \ps s may
have several \pl s in \wo s (\rd\ \& Scalzo 1997, 1998a).  
 
Calculations  indicate that it is likely that
\ml\ by \pl s in \wo s will provide an
important channel and, for low-mass (e.g., Earth-mass) \pl s, possibly the
dominant channel for \pl\ detection via \ml\ (\rd\ \& Scalzo 1997, 1998a,b
\rd\ 1998).

\section{Earth-Like Conditions}

The phrase ``Earth-like conditions" does not have a unique  meaning. 
Two requirements seem natural, however.
(1) The radiation flux received from the \cs\ should be neither too
large nor too small; (2) The \pl\ should have a rocky surface,
water, and a gaseous atmosphere.

\subsection{The Orbital Separation and the Incident Flux of Radiation}
 
For Earth-like conditions to exist, 
the primary requirement on $a,$
the orbital separation between the \pl\ and the \cs ,
 is that the incident flux of radiation from the star
should be comparable to the flux received by the Earth from the Sun.
That is, ${\cal F}/{\cal F}_\oplus$, should not be too different from unity.
\begin{equation}
{{\cal F}\over{\cal F}_\oplus}={{0.012}\over{\mu^2}}
\left( {{L}\over{L_\odot}} \right)
\, \left( {{M_\odot}\over{M}} \right) \,
\, \left( {{10\, kpc}\over{D_S}} \right)
\, \left( {{1}\over{x\, (x-1)}} \right) \simeq 1. 
\end{equation}
The mass and luminosity of the central star are
$M$ and $L$, respectively. As usual, $D_S$ ($D_L$) is the distance to the 
lensed source (lens), and $x=D_L/D_S$. 

Given that (1) the conditions
that lead to life may be flexible, (2) the effects of radiation incident
from the star are likely to be strongly influenced (either enhanced or
diminished) by the \pl 's atmosphere, and (3) internal heating
from geological processes or radioactive materials may be important,
it is not clear how large a range of values of ${\cal F}/{\cal F}_\oplus$
may be compatible with the development of life. We will therefore simply
use ${\cal F}/{\cal F}_\oplus\simeq 1$ as a guideline, and 
emphasize that this should not be viewed as an
absolute requirement.

\begin{figure}
\vspace{-1 true in}
\plotone{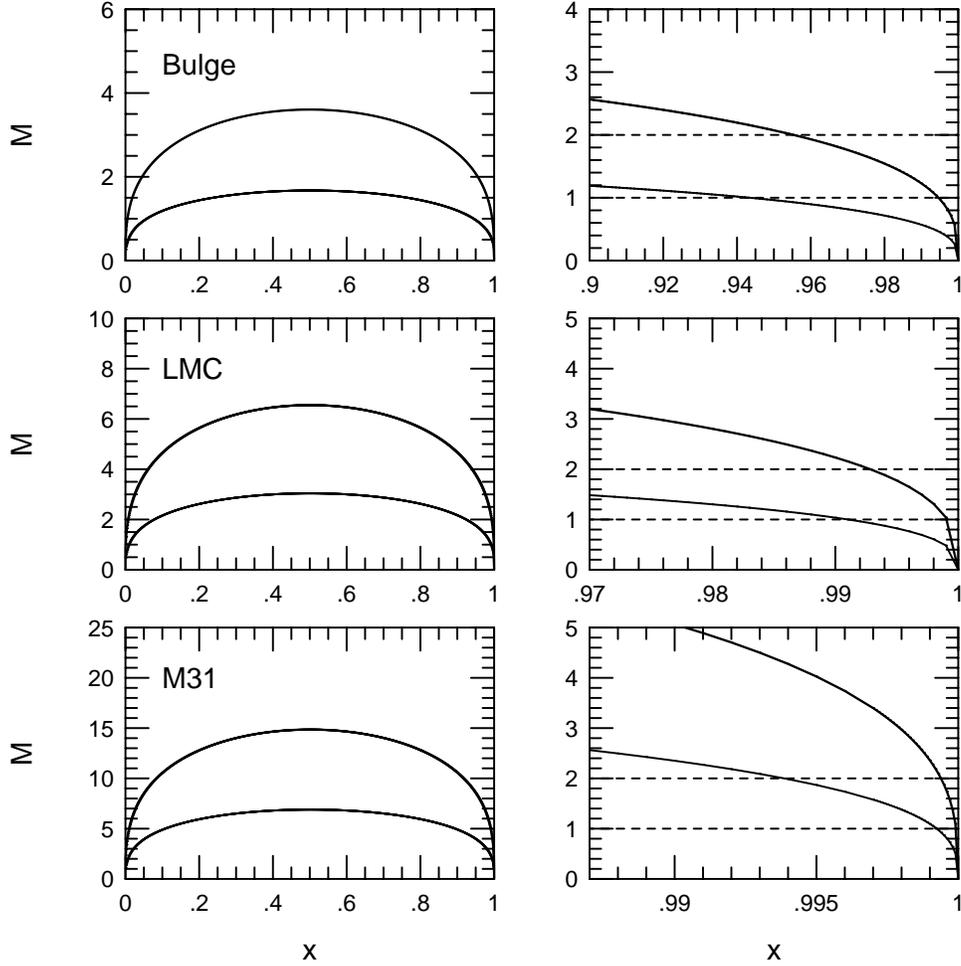}
\vspace{-2.0 true in}
\caption{Plotted is $M$ vs $x,$ where $M$ is the mass of the central star 
of the lens system and
$x=D_L/D_S.$ We have assumed that there is a \pl\ located at 
$a=\mu\, R_E,$ with $\mu=1.5$,
 and have imposed the condition: ${\cal F}/{\cal F}_\oplus=1.$
Upper curve in each panel: $L=M^{3.5}.$ Lower curve in each panel: 
$L=10 M^{3.5}.$ For the Bulge [Magellanic Clouds, M31], we have assumed
$D_S= 10$ kpc [$60$ kpc, $700$ kpc]. 
The plot in the upper left panel is for the Galactic Bulge;
the upper right panel is a blow-up of the region in $x$ that corresponds to
the lens, as well as the lensed source, being located in the Bulge.
Horizontal dotted lines are drawn at $M=M_\odot$ and $M=2\, M_\odot$ 
The left-right pairs in the middle and bottom panels show the same
quantities for the Magellanic Clouds and M31, respectively.}
\end{figure}

Figure 1 shows the relationship between $M$ and $x$ for those systems
that satisfy the relationship ${\cal F}/{\cal F}_\oplus=1.$ We have 
set $\mu = 1.5$ and have assumed
that $L=M^{3.5}$ in the upper plot, and $L=10\, M^{3.5}$ in the bottom plot;
the former would be appropriate for main-sequence stars, while 
the latter would be appropriate for slightly evolved stars.
As Eq.\ 1 and Figure 1 make clear, if the planets we will discover via
microlensing are to have incident flux comparable to the flux
incident on Earth, their stars will generally 
(although not necessarily) 
be more luminous than our sun.
This has two obvious implications. The first is that the length of
time during which the planet would have this flux incident will be
shorter than the time to date that the Earth has had roughly this flux incident.
This is because the system's star may need to be more massive than the
Sun, or even slightly evolved. The time elapsed from the
formation of the star until the present time could range from
less than $0.1$ the present age of
the sun to times comparable to the sun's main-sequence lifetime.
We do not know how long it takes for complex life forms
to develop, but it may be that the process is fast
enough that intelligent life can develop and thrive during a time
significantly shorter than the present age of the Sun.
Indeed, it is likely that a long sequence of independent processes must
occur in order for intelligent life to develop; thus, the probability
distribution may well be log normal, and the likelihood of intelligent life
developing in times much shorter than the time apparently
taken
on Earth may be significant.

The second implication is that, since the central star must be
fairly luminous, it may contribute a non-negligible fraction
of the light incident along the line of sight to the lensed source; i.e.,
the light we receive may be strongly blended. 

\subsection{The Mass of the Planet}

We may also argue, as follows, that planets likely to be deemed
Earth-like have masses within a factor of $\sim 15$ of the mass
of the Earth.  Assuming that we would like a
rocky surface, we also assume that the planet's average density should be
similar to that of Earth.  This means that the acceleration due to gravity,
$g$, scales as the cubed root of the planet's mass, $m$.  Thus, $g$ will be
within a factor of 2.5 of $g_\oplus$ (= 9.8 m/s$^2$) if $m$ is within a factor
of 15 of $m_\oplus$.  Increasing or decreasing the value of $g$ will lead to
different atmospheric contents; for any given atmospheric temperature, there
is a lower limit to $m$ (hence $g$), below which an atmosphere will not be
retained. 

Additional (and more sophisticated)
 considerations can influence these limits. For example, the
level of geothermal (planetary-thermal) activity may be less for
planets of smaller mass, influencing atmospheric chemistry.
In fact, it has been conjectured that part of the difference between Earth
and Mars, which is roughly $10$ times less massive than Earth,
 could be related to differences in planetary-thermal
activity related to their mass difference. But it is not clear that
this is the crucial difference with regard to liquid
water, for example, and other possibilities exist.
(See, e.g., 
Rampino \& Caldeira 1994.) 


In this paper, however, I will not highlight such
complementary restrictions.
This is because the main relevance of such considerations
to \ml\ observations is simply that there is likely to be a range of planet 
masses, possibly spanning two orders of magnitude,
 consistent with Earth-like conditions.
This means that
\ev\ durations for Earth-like \pl s are likely to also have a range.
The factor of $15$ derived above corresponds to a range 
of event durations from less than an hour to more than a day.
In addition, finite-source-size effects could increase the time
duration of events by a 
factor of a few (\rd\ \& Scalzo 1997, 1998a, \rd\ 1998).

\section{Exploring the Consequences: The Effects of Blending}

If the central star of a \ps\ that serves as a lens is fairly
luminous, then its light will blend with that from the
lensed star. This blending can have two consequences. First, 
if is is measurable and can be quantified, blending
makes it possible to  
learn more about the lens system: the spectral type of the 
\cs\ can be determined and, in some cases, even the mass of the \pl\
may be inferred (\rd\ 1998). Unfortunately, however, the second consequence of 
blending is that the peak magnification, or in some cases,
 the time interval
during which the \ev\ is observable, may be decreased to the point that
the \pl-lens  \ev\ is not detected at all.  
We must distinguish between the  detection of \pl s in the
\zres, and the detection of \pl s in wider orbits. 
The discussion below applies to cases in which the orbital separation
is outside the \zres, when the \pl\ acts as a more-or-less
independent lens. At the end of \S 4.1 we return to the case of \pl s in
the \zres . 

Let $f$ represent the fraction of the baseline flux contributed
by the lensed star.
When the central star of the \ps\
 is luminous, particularly if its flux comes close to
satisfying the criterion studied in the last section
(${\cal F}/{\cal F}_\oplus \sim 1$), then $f$ can indeed be small enough for the
effects of blending to be measurable.
If, for example, the apparent $V$ magnitude of the combined light
coming along the line of sight from a lensing \ev\ is [$19.5, 17.0,
15.7$], then $f\leq 0.1$ if the lens is 
located in the Bulge and is a main-sequence star of mass
roughly equal to [$1, 2, 3$] $M_\odot$.
A small value of $f$ can allow us to reliably determine
the effects of blending and to thereby learn more about the lens.
The question we address below is whether values of $f$ small enough to  
be useful may prevent the \ev\ from being detected.
\footnote{In addition to knowing whether events will be detectable,
one would also like to know whether or not it will be possible
to extract definitive evidence of blending. Studying the \lc\
alone provides limited information. Distant approaches (\rd\ \& Esin 1995,
Wo\'zniak \& Paczy\'nski 1997) lead to \lc s that are difficult to
"deblend" using the \lc\ alone. So does blending with $f<<1$
(Wo\'zniak \& Paczy\'nski 1997).
In the latter case, however, evidence of blending can be obtained
through comparisons of spectra taken at peak with spectra taken
at baseline (unless the lens and source are of similar spectral type).    
Astrometry can also be useful 
(see, e.g., Goldberg 1988, Goldberg \& Wozniak 1998).}  

\subsection{The Effects of Blending on Event Detection}

\begin{figure}
\plotone{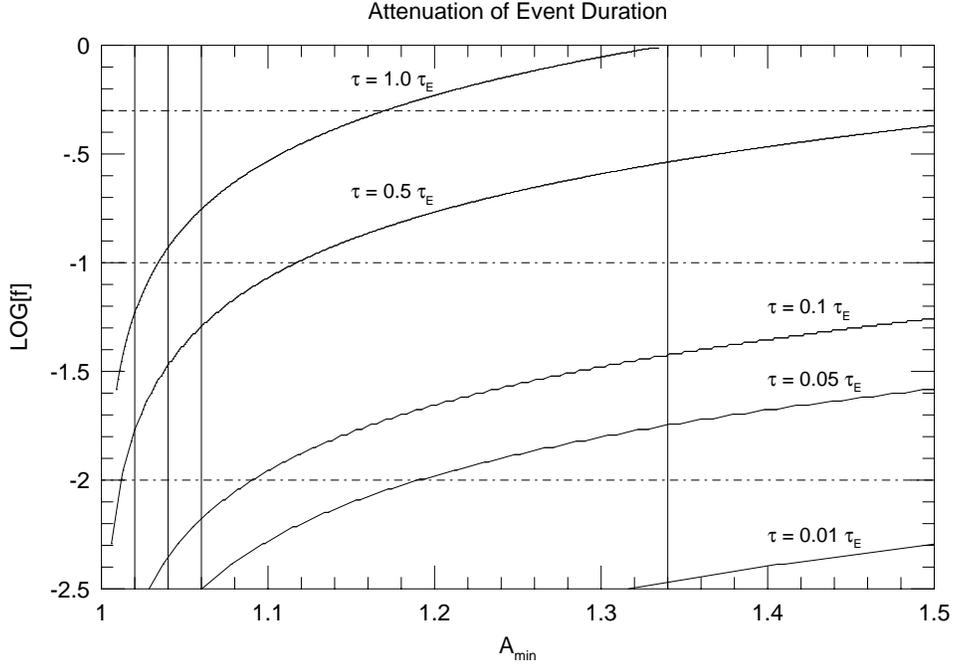}
\vspace {-4 true in}
\caption{
Log[f] vs $A_{min}$. 
Source tracks going straight through an Einstein diameter have been
considered. 
$\tau_E$ is the time needed for the source to cross a distance
equal to the Einstein diameter.  
The observed \ev\ \du , $\tau,$ is defined to be the time during which
$A>A_{min}$.  
Because each \ev\ was subject to blending ($f < 1$),
$\tau$
 is attenuated relative to the event duration
without blending. 
Note that, for $A_{min} < 1.34,$ the duration of the observed \ev\ would be
greater than $\tau_E$, were there no blending.
Each curve shown corresponds to a fixed ratio $\tau/\tau_E.$
The horizontal lines correspond to $f_V = 0.5, 0.1, 0.01,$ proceeding
from top to bottom. For reference, Note that, if the lens is a main-sequence
star with mass [$1, 2, 3$] $M_\odot,$ then $f_V=0.5$ if the total
apparent V magnitude along the line of sight is [$20.2, 17.6, 16.4$];
similarly, $f_V=0.01,$ for $M_V$ = [$19.4, 16.9, 15.6$].       
Vertical lines correspond to $A_{min}=1.02, 1.04, 1.06;$ these 
provide comparisons which  
demonstrates the improvement in \ev\ detection relative to the case $A=1.34$,
which is also marked by a vertical line.
}
\end{figure}

When light from the lensed source is blended with light from other
sources, the observed \mage , $A_{obs},$ is smaller than the true
\mage , $A.$ 
\begin{equation} 
(A_{obs}-1)=f\, (A-1) 
\end{equation} 
Thus, in order for a \lc\ perturbation to be brought above the 
detection limit, $A$ must be larger than it would otherwise have to
be, the projected distance between the source and lens must be smaller,
and the \ev\ will consequently appear to have a shorter \du .
To describe this effect systematically, \rd\ \& Esin (1995)
introduced the ``blended Einstein radius", $R_{E,b}$.  
Let $A_{min}$ be the minimum peak \mage\ needed for \ev\ detection.
Define the \du\ of each \ev\ to be time during which the \mage\
was greater than $A_{min}.$ The expression for the blended Einstein radius
is then  
\begin{equation}
R_{E,b}= R_E
\sqrt{2}\sqrt{{{(A_{min}-1)+f}\over{\sqrt{(A_{min}-1)^2 
+ 2 (A_{min}-1)\ f}}}-1}.
\end{equation}
Figure 2 illustrates the influence of blending on \ev\ \du\
as a function of both $A_{min}$ and $f.$ 
For a given value of $f,$ the \ev\ \du\ is longer if $A_{min}$ is
smaller. Thus, increasing the photometric sensitivity to accomodate
smaller values of $A_{min}$
should increase the detection rate.
Note that if $A_{min}$ is $1.06,$ then even if $f=0.18,$ the
\ev\ \du\ (which would have been $2\, \tau_E$), is reduced by only a factor
of $2,$ to $1\, \tau_E.$ 
Thus, while blending does tend to decrease the \ev\ \du, making more 
frequent \mo\ desirable, we can expect to be able to
detect a large majority of \ev s by using sensitive photometry.
Note, however, that, even with $A_{min}=1.34,$ the time \du\ of
an \ev\ with $f=0.32$ would be decreased by only a factor of $2$.  
If $A_{min}$ could be reduced to $1.02$--a formidable
task for a large-scale \mo\ program--then for $f=0.03,$
an \ev\ that would have lasted for $3\, \tau_E$ will have an observed
\du\ of $1\, \tau_E.$   


There has not been a detailed study of the effects of blending on the
detection of \pl s in the \zres . We can make some general observations,
however. 
Blending does decrease the duration of stellar-lens \ev s. Since
we expect most such \ev s to last for times on the order of weeks
or months, the arguments above show that it is unlikely that
blending will cause us to miss the stellar-lens \ev\ altogether.
What blending can do,  though, is to shorten the \ev\ in such a way that
we become aware of it only after the \pl-lens perturbation
has occurred, if the perturbation takes place early in the event.  
Perturbations can take place early in stellar-lens events (Paczy\'nski
1996, Wambsganss 1997~\footnote{Note that Wambsganss refers to a thin annulus
as the ``resonant zone". This annulus is
just a small part of what we refer to as the ``\zres".
In fact, in his comprehensive study, Wambsganss has effectively mapped out
the extent of the \zres. For the mass ratios he considers it
extends from $\sim 0.6 \, R_E$ to $\sim 1.6\, R_E$.}). 
Thus, one
effect of blending is to decrease 
the rate of observable resonant \pl -lens \ev s. The effect should
not be large, however, because (1) the \ev s that occur early
are not generally the most distinctive and readily observed \pl-lens
\ev s, and   
(2) careful monitoring,
once we know the stellar-lens \ev\ is underway, can help us to catch
even perturbations that take place as the 
measured flux declines toward baseline, 
so that we suffer the worst losses only on the upswing of the \mage .
The second effect of \bl\
is to alter the \lc\ shape. When, however, \fsse\
are unimportant, the spike in \mage\ associated with
resonant events is so distinctive that frequent monitoring
of the \lc\ during the event should allow us to detect it.    
The remaining challenge is therefore 
to estimate the combined effects of finite-source size and blending
on the overall efficiency of detecting \pl s in the \zres.
 
\subsection{The Effects of Blending and Finite Source Size on Studying the Planetary System}

The up-side of blending is that the quantity and
color of light from the \cs\ are themselves valuable pieces of 
information about the \ps . Thus, if the blending is
significant enough to make these features
measurable,  it allows us to learn something interesting
about the \ps . 
We would like to do the following.

\noindent (1) Definitively establish that there is blending.

\noindent (2) Establish that the light not emanating from the
lensed source is most likely emitted by the \cs\ of the \ps .   
This establishes that, even if the \ev\ was an isolated
\ev\ of \shdn, the lens was a \ps .
 
\noindent (3) Determine the spectral type, and possibly the
mass of the \cs . In cases in which the mass ratio can be extracted
from the \lc , this will establish the mass of the planet lens.

If, in addition to blending, there are measurable \fsse , then 
the mass of any \pl s that served as lenses may be determined directly.

These possibilities, together with the combined effects
of blending and \fsse\ on \ev\ detection,
 are discussed in more detail elsewhere (\rd\ 1998), since they
apply to all \ps s discovered via \ml, not just those most likely
to include an Earth-like \pl . Here I simply note that
a combination of observations  would generally be needed 
to accomplish these goals. These include: \lc\ studies to
begin to assess the role of blending and \fsse ; possible astrometric
studies (e.g., Goldberg 1988, Goldberg \& Wozniak 1998,
Boden, Shao, \& Van Buren 1998, Dominik \& Sahu 1998, Mao \& Witt 1998);    
spectra taken near peak \mage, compared with spectra taken at baseline,
to determine the spectral type and radius of the lensed star;
and high-spatial-resolution follow-up observations to determine
whether any light not emanating from the lensed star can be explained
by a chance superposition of light from other stars in the field.

\section{Conclusion}

Microlensing searches for \pl s complement other types of searches.
They have the advantage of probing vast volumes of space, and of
providing information about the existence of \pl s in diverse
stellar environments. An additional advantage is that \ml\
can discover low-velocity \pl s.
Because they are far away, however,
\pl s discovered via \ml\ are less amenable to detailed follow-up
observations. In this paper I have shown that \ml\ is a tool that can
discover distant \pl s with Earth-like conditions, should such
\pl s exist. In addition, 
I find  that, among \pl s discovered by \ml\ surveys,
 those likely to be experiencing Earth-like conditions are 
particularly good targets for some follow-up studies.
 This is because the \cs\ of these 
\ps s may be luminous enough that the blending of its light
with light from the lensed source may be detectable and quantifiable.
Through a combination of spectra taken during the \ev,
high resolution images and spectra taken after the \ev ,
and \lc\ fitting, we may be able to determine the spectral 
classification of the \cs , the distance of the \ps\ from us,
and even, in some cases, the mass of the \pl\ lens. (See also \rd\ 1998.) 
Although there may be a subset of such \ev s that are rendered
undetectable by the very blending that helps to make them
potentially so interesting, the results presented here show that
in most cases, the \pl-lens \ev s should be detectable.  

Until now, \ml\ discussions of Earth-like \pl s have intrinsically
focused on Earth-mass plants. In addition, because (1) low-mass
stellar lenses are almost certainly significantly more numerous than
higher-mass stars, and (2) a larger value of the mass ratio between the
\pl\ and star is preferred for detection, it has become
common for calculations to focus on a low-mass ($\sim 0.3\, M_\odot$)
star orbited by an Earth-mass
\pl . Figure 1 indicates, however, that it is unlikely that a low mass
star 
will harbor a \pl\ with Earth-like conditions that can be
discovered via \ml . Furthermore, the considerations in \S 3.2
indicate that there may be room for  more flexibility in the mass of \pl s
with Earth-like conditions than has been assumed so far.    
Simple arguments indicate
that the range
could extend as much as a factor of ten above, and a factor of 10
below the mass of the Earth.

Finally, it is important to note that
we should consider the possibility that \ml\ (whether by resonant
or wide \pl s) is most likely to discover the outer \pl s in a
system that may contain closer planets experiencing Earth-like
conditions. In this case, the \ml\ \ev s serve as beacons directing
us to the \ps . In the near-term, such
discoveries can contribute to developing the statistics of distant \ps s,
such as the frequency of \pl s as a function of spectral type and the
spatial distribution of \pl s around stars. In the far (and so
far unforeseeable) future,
astronomers with instruments capable of measuring the spectra of distant
stars and detecting small-amplitude Doppler effects can perhaps
check these \ps s for the presence of \pl s in  orbits smaller than
those occupied by the \pl s discovered via \ml .

\bigskip
\bigskip
\bigskip
\centerline{ACKNOWLEDGEMENTS}

It is a pleasure to thank the referee, Scott Gaudi, for suggested changes to
an earlier manuscript (\rd\ \& Scalzo 1998b) that contained much of this
work; his suggestions
have helped to
improve the presentation.
I would like to thank 
Arlin Crotts, Andrew Gould, Jean Kaplan, Christopher Kochanek, 
David W. Latham, Avi Loeb,   
Shude Mao, Robert W. Noyes, Bodhan Paczy\'nski,
Bill Press, 
Penny Sackett, Kailash Sahu,  
Michael M. Shara, Edwin L. Turner,  Michael S. Turner,  
and the participants in the 1997 Aspen workshops, ``The Formation
and Evolution of Planets" and ``Microlensing"
for interesting
discussions. It is also a pleasure
to thank the Aspen Center for Physics and the Institute for
Theoretical Physics at Santa Barbarba for their hospitality while
the first version of this and related papers was being written,
and the Inter-University Center for Astronomy and Astrophysics in Pune,
India for its hospitality while the paper was revised.
This work was supported in part by NSF under GER-9450087 and
AST-9619516, as well as by funding from AXAF.

{}

\clearpage

\begin{thebibliography}{}









\bibitem[Angel \& Woolf 1997]{aw97}
Angel, J. R. P., \& Woolf, N. J. 1997, \apj, 475, 373









\bibitem[Boden, Shao, \& Van Buren 1998]{bsvb98}
 Boden, A.F., Shao, M., \& Van Buren, D. 1998, astro-ph/9802179






\bibitem[Cochran {\it et al.} 1997]{coch97}
Cochran, W.D., Hatzes, A.P., Butler, R.P., \& Marcy, G.W. 1997, ApJ, 483, 457 





\bibitem[Di Stefano, R.  \& Scalzo, R.A. 1998a]{rdras98a}
Di Stefano, R. \& Scalzo, R.A. 1998a,  submitted to ApJ. 
 
\bibitem[Di Stefano, R.  \& Scalzo, R.A. 1998b]{rdras98b}
Di Stefano, R. submitted to ApJ. 
 
\bibitem[Di Stefano, R.  \& Scalzo, R.A. 1997]{rdras97}
Di Stefano, R. \& Scalzo, R.A. 1998c,  astro-ph/9711013. 
 
\bibitem[Di Stefano 1998]{rd98}
Di Stefano, R. 1998,  in preparation. 
 
\bibitem[Di Stefano \& Mao 1996]{rdmao96}
Di Stefano, R., \& Mao, S.  1996, ApJ, 457, 93
 
\bibitem[Di Stefano \& Esin 1995]{rdmao96}
Di Stefano, R., \& Esin, A. A.  1995, ApJ, 448, L1

\bibitem[Dominik \& Sahu 1998]{domsahu98}
Dominik,M. \& Kailash, K.C. 1998, astro-ph/9805360

\bibitem[Goldberg 1998]{gold98}
Goldberg, D.M. 1998, ApJ, 498, 156

\bibitem[Goldberg \& Wozniak 1998]{goldetal98}
Goldberg, D.M., Wozniak, P.R. 1998, AcA, 48, 19   

\bibitem[Feinberg \& Shapiro 1980]{feinb97}
Feinberg, G. \& Shapiro, R. 1980, ``Life Beyond Earth",
William Morrow \& Co., NY





\bibitem[Gould \& Loeb 1992]{loeb92}
Gould, A. \& Loeb, A.  1992, \apj, 396, 104





\bibitem[Kamionkowski 1995]{kski95}
Kamionkowski, M.  1995, ApJ, 442, L9



\bibitem[Labeyrie 1996]{lab96}
Labeyrie, A.  1996, A\&AS, 118, 517L




\bibitem[Marcy {\it et al.} 1997a]{marc97a}    
Marcy, G.W. {\it et al.} 1997, ApJ, 481, 926.  

\bibitem[Marcy {\it et al.} 1997b]{marc97b}    
Marcy, G.W. {\it et al.} 1997, ApJL, 474, 115.  

\bibitem[Butler \& Marcy 1996]{butmar96}
Butler, R.P., \& Marcy, G.W. 1996, ApJ, 464L, 153 

\bibitem[Marcy \& Butler 1996]{butmar96}
Marcy, G.W., \& Butler, R.P. 1996, ApJ, 464L, 153 

\bibitem[Mayor \& Queloz 1995]{may95}
Mayor, M., \& Queloz, D. 1995, Nature,
378, 355 


\bibitem[Mao \& Paczy\'nski 1991]{mao91}
Mao, S. \& Paczy\'nski, B.  1991, \apj, 374, L37

\bibitem[Mao \& Witt 1998]{maowitt98}
Mao, S. \& Witt, H.J.  1998, astro-ph/9804045 

\bibitem[McCord et al. 1997]{mccord97}
McCord, T.B. et al. 1997, Science, 278     

\bibitem[Noyes {\it et al. 1997}]{noyes97}
Noyes, R.W. 1997, ApJ, 483L, 111


%
\bibitem[Paczy\'nski  1996]{[pacz96}
Paczy\'nski 1996, ARA\&A, 34, 419
 

\bibitem[Peale 1997]{peale97}
Peale, S. J. 1997, Icar, 127, 269
 



\bibitem[Rampino 1994]{rampino94}  
Rampino, M.R.  \& Caldeira. K 1994, ARAA, 32, 83. 

\bibitem[Raulin et al. 1992]{raul92}
Raulin, F., et al. 1992,  In ESA, Symposium on Titan,  149 


\bibitem[Reynolds et al. 1983]{rey83}
Reynolds, R. T., et al.  1983 Icarus, vol. 56, 246

\bibitem[Sagan, Thompson, \& Khare 1992]{sagan92}
Sagan, C., Thompson, W. R., \& Khare, B. N. 1992, In ESA, Symposium on Titan,
    161

 





 


\bibitem[Wambsganss 199]{wamb97}
Wambsganss, J.  1997, MNRAS, 284, 172

\bibitem[williams et al 1997]{willkastwade97}
Williams, D.M., Kasting, J.F., \& Wade, R.A. 1997, Nature, 385, 234




\end{thebibliography}
\end{document}